\pdfoutput=1 
\RequirePackage{ifpdf}
\documentclass{JINST}
\usepackage{subfigure}
\usepackage{url} 

\title{Design and test of an extremely high resolution Timing Counter for the 
MEG II experiment: preliminary results }

\author{M.~De Gerone$^a$\thanks{Corresponding author.}, F.~Gatti$^{a,b}$, W. Ootani$^c$, Y.~Uchiyama$^{c}$, M.~Nishimura$^c$, S.~Shirabe$^d$, P.W.~Cattaneo$^e$, M.~Rossella$^e$\\
\llap{$^a$}Istituto Nazionale di Fisica Nucleare, Sezione di Genova,\\
  Via Dodecaneso 33, 16146, Genova (GE), Italy\\
\llap{$^b$}Universit\'a degli Studi di Genova,\\
  Via Dodecaneso 33, 16146, Genova, Italy,\\
\llap{$^c$}International Center for Elementary Particle Physics, University of Tokyo\\
 7-3-1 Hongo, Bunkyo-ku, Tokyo 113-0033, Japan\\
\llap{$^d$}Department of physics, Kyushu University\\
 6-10-1 Hakozaki, Higashi-ku, Fukuoka 812-8581, Japan\\
\llap{$^e$}Istituto Nazionale di Fisica Nucleare, Sezione di Pavia,\\
 Via Agostino Bassi, 6, 27100, Pavia (PV), Italy\\
E-mail: \email{degerone@ge.infn.it}}

\abstract{The design and tests of Timing Counter elements for the upgrade of the MEG experiment, MEG II,
are presented. The detector is based on several 
small plates of scintillator with a Silicon PhotoMultipliers dual-side readout. The optimisation of 
the single counter elements (SiPMs, scintillators, geometry) is described. Moreover, the results 
obtained with a first prototype tested at the Beam Test Facility (BTF) of the INFN Laboratori Nazionali 
di Frascati (LNF) are presented.}


\begin{document}

\section{Introduction: the MEG experiment}\label{sec:intro}

The MEG experiment has been running since 2008 at the Paul Scherrer Institut (Villigen, CH), looking for the $\mu\rightarrow e\gamma$ decay. The MEG collaboration recently published the results based on the analysis of data collected in the years 2009-2011: BR($\mu\rightarrow e\gamma$)$\le 5.7\times 10^{-13}$ @90$\%$ C.L. \cite{bib:MEG2013}. While the analysis of the 2012-2013 data is still ongoing, an upgrade program of the MEG experiment (MEG II) has unfolded since 2012 \cite{bib:MEGup}, aiming to improve the experiment sensitivity by an order of magnitude, down to $\sim 5\times 10^{-14}$. In order to reach such a sensitivity, most of the current detectors have to be re-designed or modified. In this paper, the development of an extremely high resolution detector for the measurement of the positron timing is described in detail.

\section{The Timing Counter upgrade}\label{sec:TCup}

\begin{figure}[hc] 
\centering
   \includegraphics[width=0.9\textwidth]{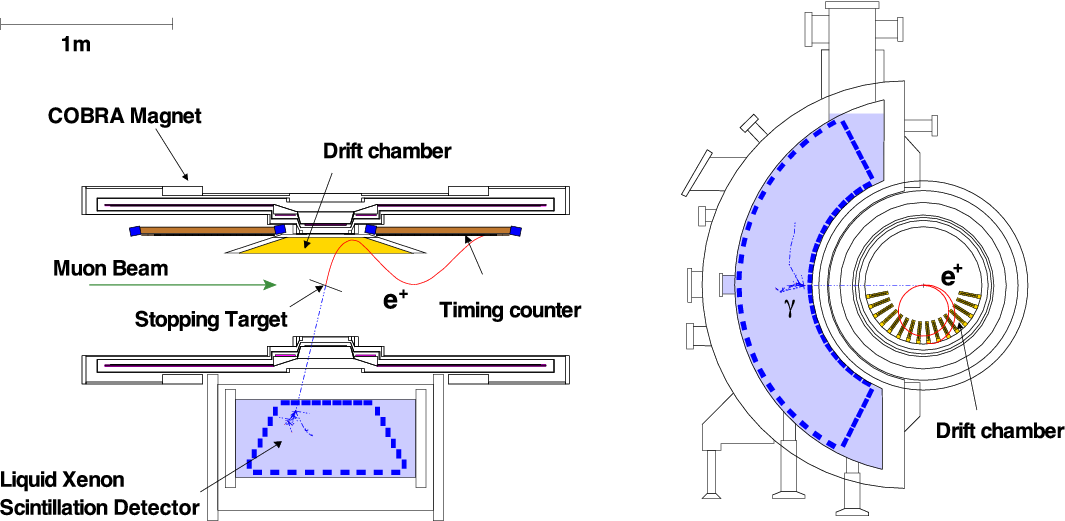}
   \caption{Schematic view of the MEG detector: side and front views.}
\label{fig:megold}
\end{figure}

The MEG detector \cite{bib:MEGdetector} is designed to measure with the highest possible resolution the kinematic variables that define the signature of the decay $\mu\rightarrow e\gamma$. Photons are detected by a Liquid Xenon detector placed outside the magnetic spectrometer where positrons are reconstructed (see figure \ref{fig:megold}). The spectrometer is made of a superconductive magnet, a set of Drift CHambers (DCH) and the Timing Counter (TC). The DCH system, together with the specially designed field provided by the COBRA magnet, measures positron energy and emission angle, while the purpose of the TC is to measure the positron time of impact.

\begin{figure}[ht]
  \centering
   \includegraphics[width=0.7\textwidth]{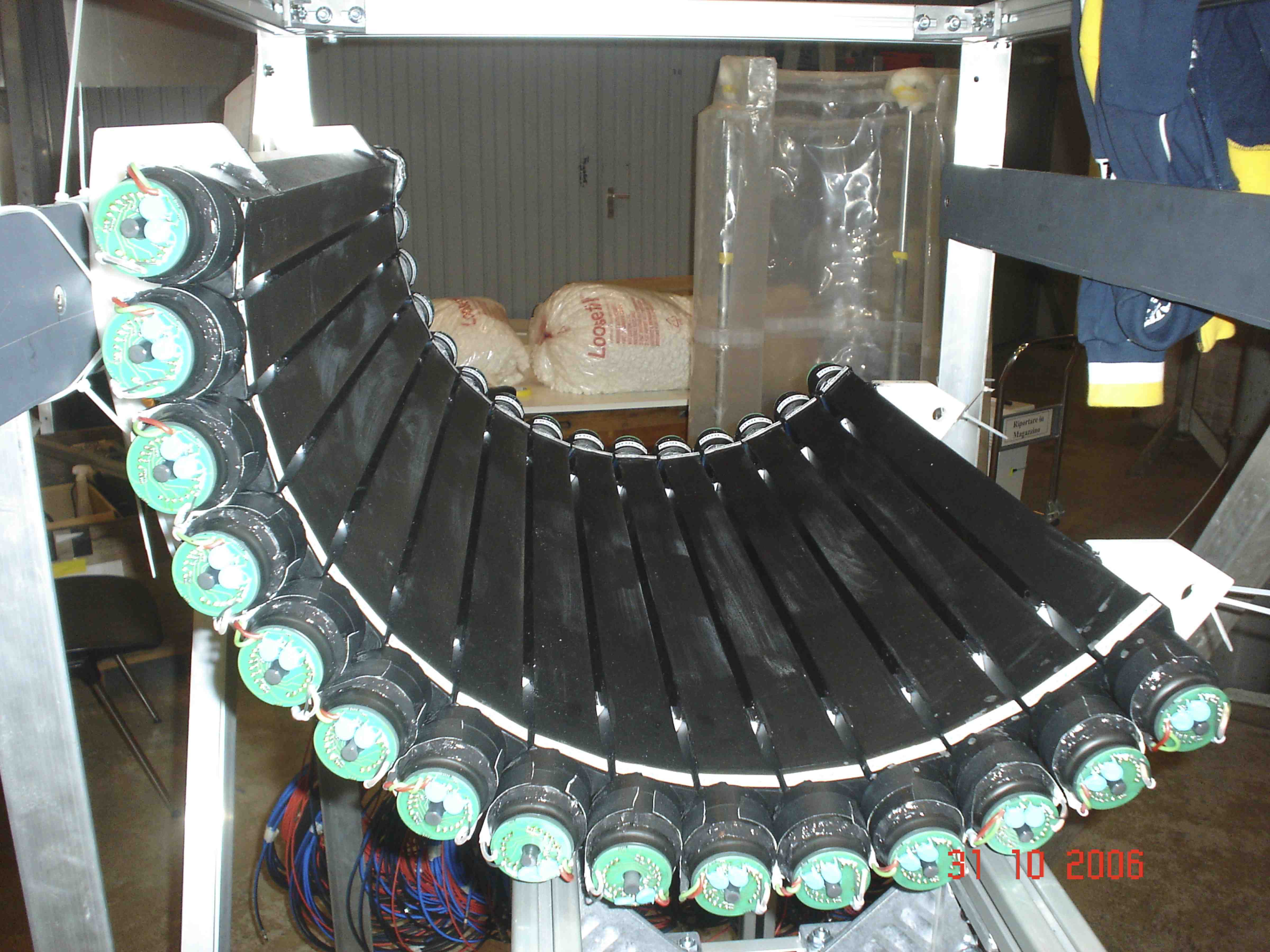}
   \caption{Picture of the current Timing Counter. PMT and scintillator bars lodged in a black plastic socket are visible.}
\label{fig:currentTC}
\end{figure}

\begin{figure}[hc]
\centering
\includegraphics[width=1.\textwidth]{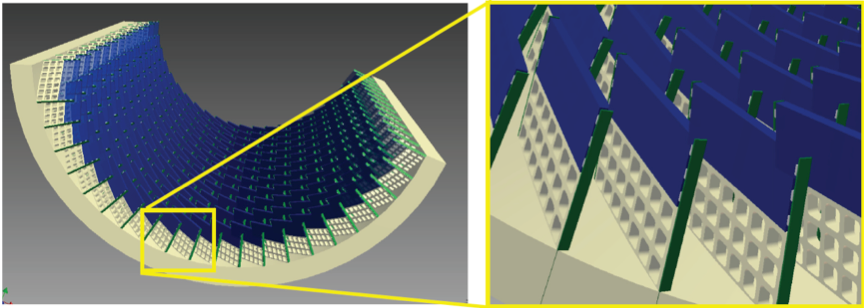}
\caption{Schematic view of the new timing counter design. On the left: overview of the detector. On the right: detail of the single counter configuration.}
\label{fig:newTC}
\end{figure}

The current Timing Counter \cite{bib:TCpaper} is made of two identical arrays (placed inside the magnet 
up- and down-stream the target position) of 15 scintillating bars (Bicron BC404), with $80\times 4 \times 4\, 
\mathrm{cm}^{3}$ size arranged in a barrel-like shape (see figure~\ref{fig:currentTC}). 
Each bar is read-out on both sides by a fine mesh PhotoMultiplier Tube (PMT, Hamamatsu R5924). 
Signals from PMTs are processed to be fed into the trigger and DAQ system. 
The Timing Counter has been running since 2008, showing good and stable time resolution of $\sim65$ ps \cite{bib:TCcalib}.\\
Some issues suggest that the design of the detector has to be changed to increase the resolution:
\begin{itemize}
\item the PMT operation in high magnetic field and helium environment deteriorates the PMT transit time spread and gain, also using fine mesh PMTs;
\item large size scintillator bars generate uncertainties on impact point reconstruction and spread of the trajectories of the optical photons inside the scintillator itself;
\item the large amount of material crossed by the positron in the TC bar prevents the use of
hits beyond the first one.
\end{itemize}
These problems originate from the usage of PMTs and large size scintillator bars. Thus, the natural choice is to 
increase the detector granularity and to upgrade the read-out system, exploiting the recent development of fast high 
gain solid state detector like Silicon PhotoMultipliers (SiPMs). 
A detector consisting of many scintillator plates (from now on: pixel) with SiPM read-out allows 
to overcome the limitations of the current TC (see figure~\ref{fig:newTC} as possible layout):
\begin{itemize}
\item magnetic field has no influence on SiPM operation;
\item higher granularity results in smaller uncertainties from impact position measurement;
\item thanks to the smaller amount of material along the positron trajectory, it is possible to take advantage of the information coming from all the pixels crossed by the particle.
\end{itemize}
The last point is quite remarkable, because the time resolution is expected to improve as $1/\sqrt{N_{hit}}$, where $N_{hit}$ is the number of pixels crossed by the positron. 
Moreover, the small size of the single element results in a more flexible configuration of the detector, allowing the possibility to tailor the position and the density of the pixels along the detector. Also, very short rise time scintillator (like Bicron BC422, see section \ref{sec:scintcomp}) can be used even in presence of a short attenuation length.
\\
Single counter good performances have already been proved \cite{bib:pixelStoykov,bib:pixelWataru}. In the following, the research and development work on several prototypes in order to choose the best material and device is presented.

\section{Single counter optimisation}\label{sec:pixelopt}

The optimisation of the single counter configuration started from a systematic study among the SiPMs and the scintillators available to compare the properties relevant for our application. 

\subsection{SiPM comparison}\label{sec:sipmcomp}

Silicon Photomultipliers are good candidate for the new TC, thanks to their 
characteristics: good time resolution, quite high gain, compactness. 
We tested different models 
from Hamamatsu Photonics, Advansid, Ketek and SensL. All these devices share some features: 
they have a size $3\times3$ mm$^3$, in order to be easily coupled to few mm thick scintillator 
pixels, and a good sensitivity in the near ultraviolet range, in order to match common plastic 
scintillators emission spectra. The SiPM models under test are summarised in table \ref{tab:SiPM}.

\begin{table}[h]
\caption{Summary of SiPMs model tested.}
\label{tab:SiPM}
\smallskip
\centering
\footnotesize
\begin{tabular}{|c|ccc|}
\hline
Manufacturer&Model&Type&Note\\
\hline
&S10362-33-050C&Conventional (Old) MPPC&Ceramic package\\ 
&S10931-050P&&surface mount\\ \cline{2-4}
Hamamatus Photonics&S12572-050C(X)&New (standard type) MPPC&Metal quench resistor\\
&S12572-020C(X)&&25 $\mu m$ pitch\\ \cline{2-4}
&S12652-050C(X)&Trench-type MPPC&Metal quench resistor\\
&3X3MM50UMLCT-B&&Improved fill factor\\
\hline
Advansid&&NUV type&\\
\hline
Ketek&PM3350 prototype-A&Trench Type&\\
\hline
SensL&MicroFB-30050-SMT&B-Type&Fast output\\
\hline
\end{tabular}
\end{table}

For each model, the noise level (dark count rate and cross-talk) together with the Photon Detection Efficiency (PDE) has been evaluated. Moreover, also the breakdown dependence on temperature has been evaluated. Finally, the timing resolution has been measured on a prototype pixel with fixed sizes.

\paragraph{Setup}
SiPMs are put in a thermal chamber, which allows to keep the device at fixed temperature (23$^\circ$C in the following measurements). Signals are transmitted on a coaxial cable to a voltage amplifier (developed at PSI, based on MAR-6SM amplifier \cite{bib:pixelStoykov}), then they are sampled at 5 Gs/s by a waveform digitiser (DRS4 evaluation board \cite{bib:DRSStefan,bib:DRSStefanNIM} also developed at PSI). 

\begin{figure}[ht!]
     \begin{center}
        \subfigure[]{%
            \label{fig:darkcount}
            \includegraphics[width=0.5\textwidth]{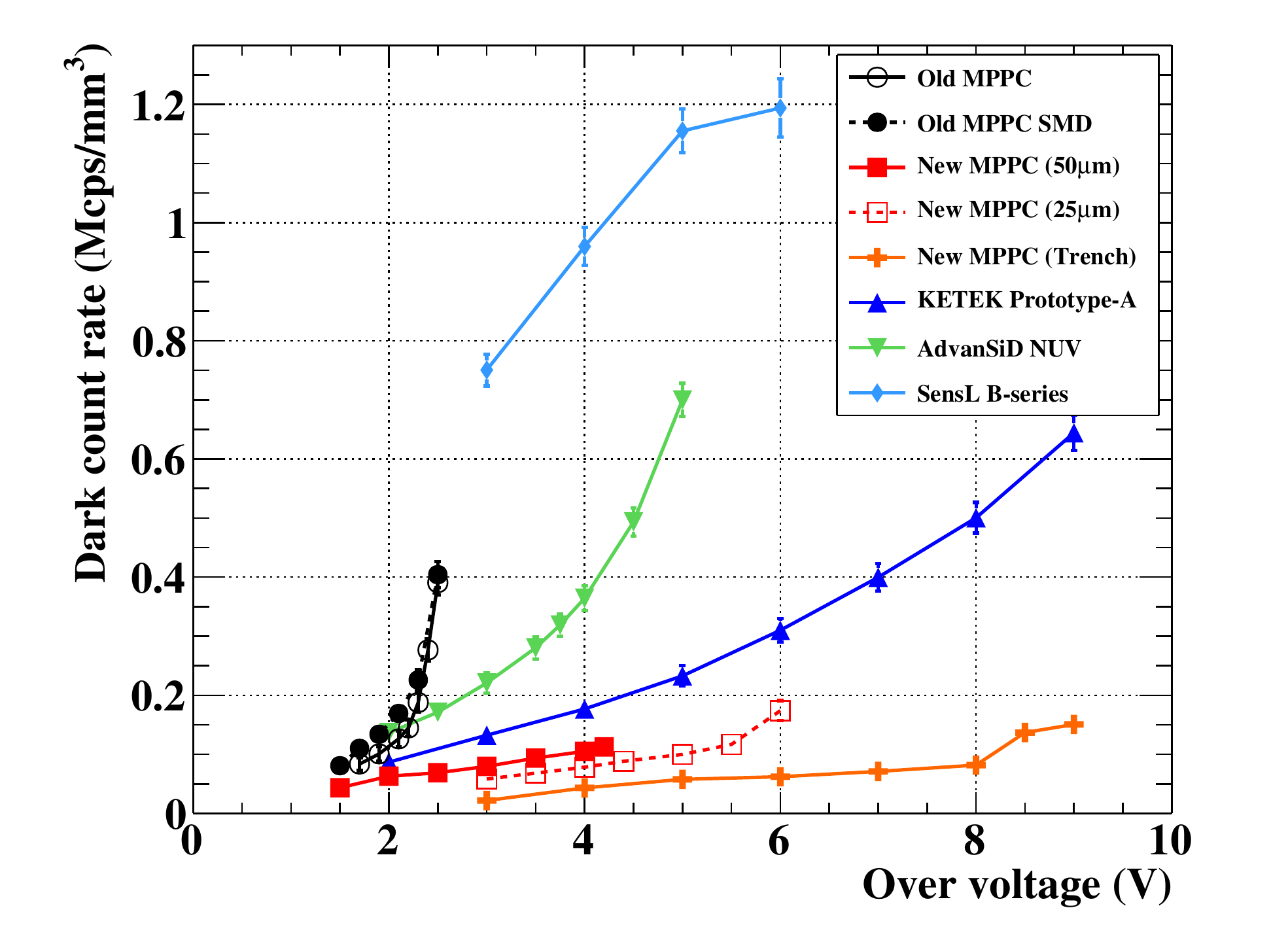}
        }%
        \subfigure[]{%
           \label{fig:crosstalk}
           \includegraphics[width=0.5\textwidth]{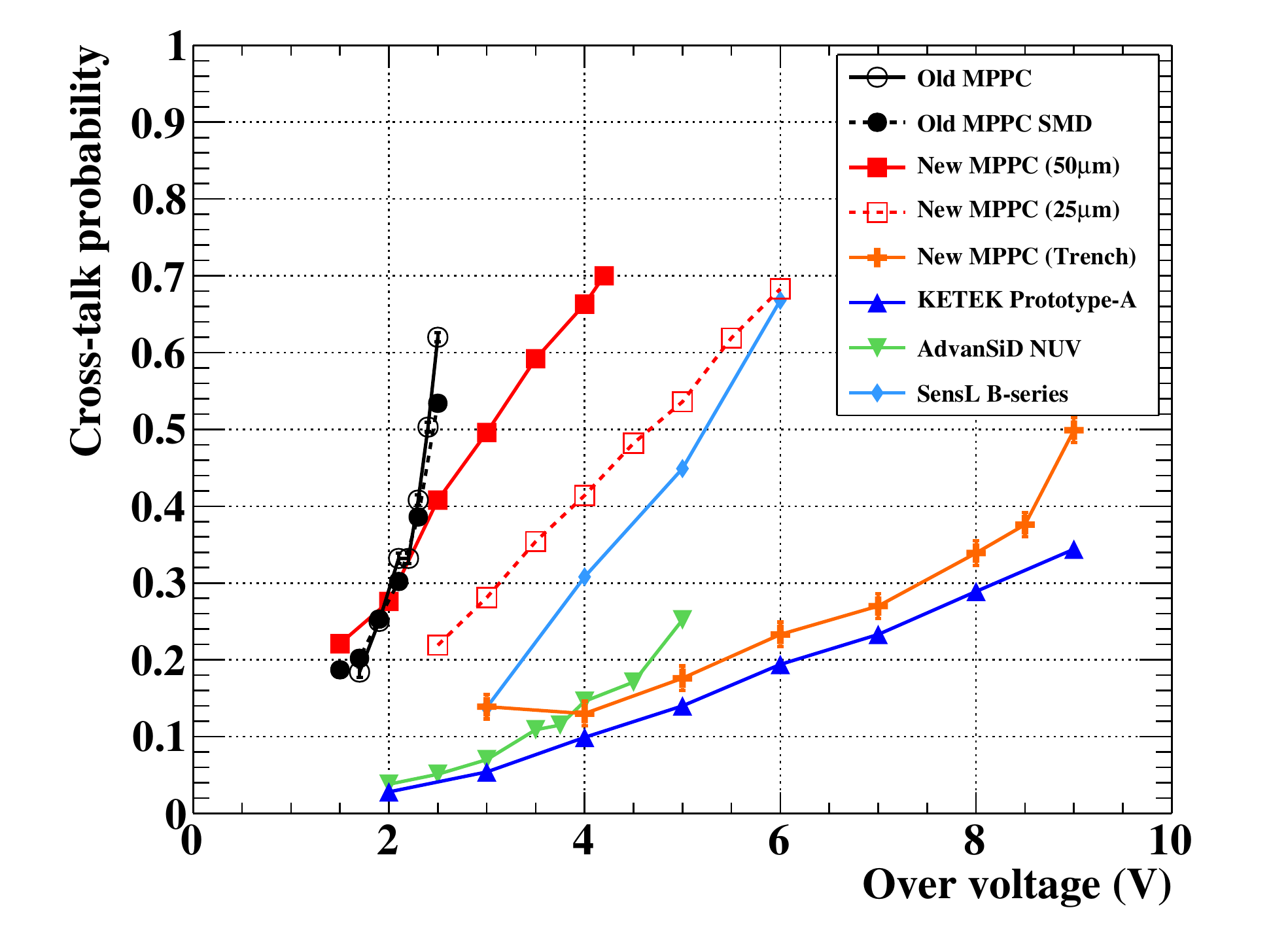}
        }\\ 
        \subfigure[]{%
            \label{fig:pde}
            \includegraphics[width=0.5\textwidth]{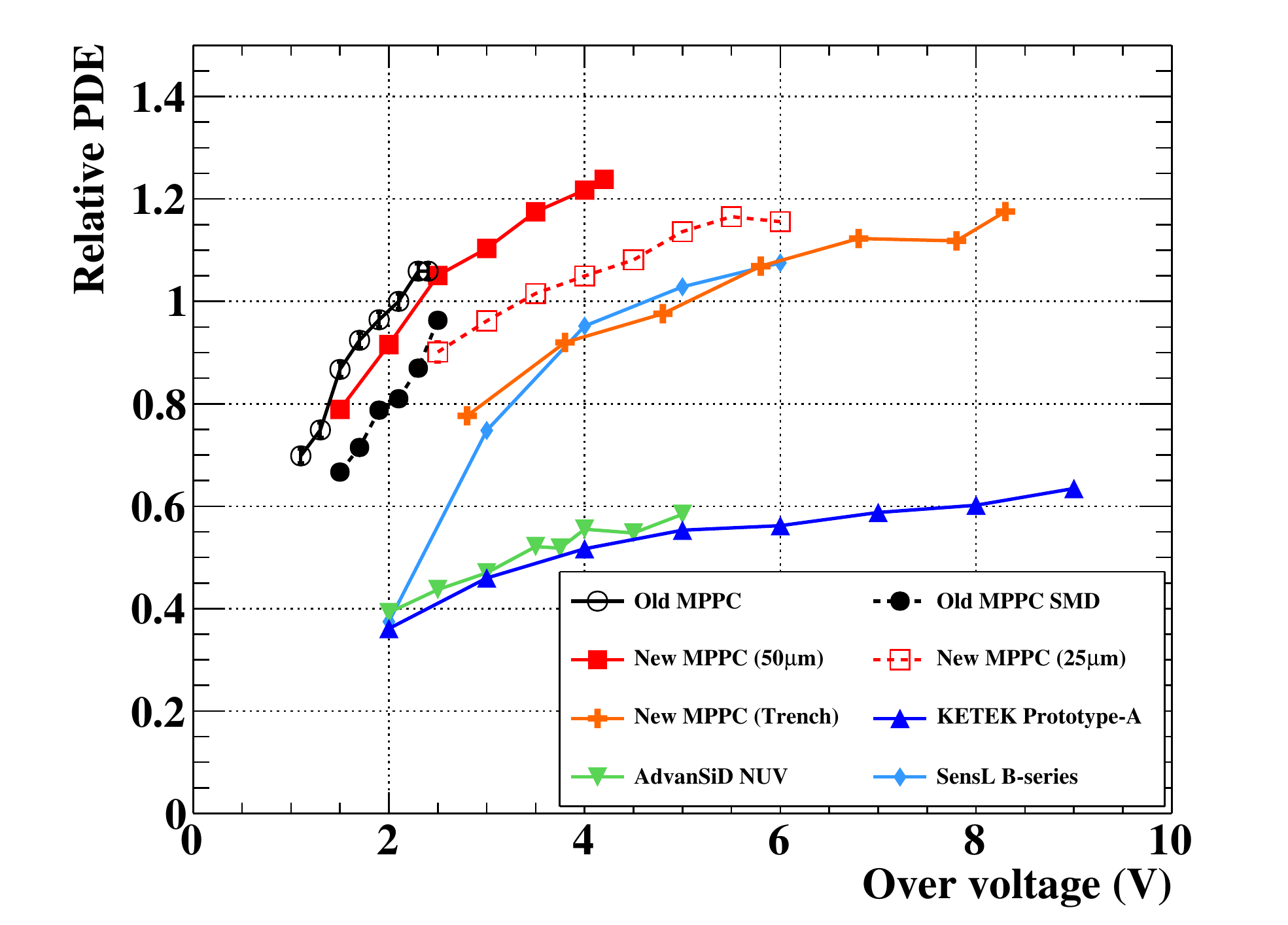}
        }%
        \subfigure[]{%
            \label{fig:timereso}
            \includegraphics[width=0.5\textwidth]{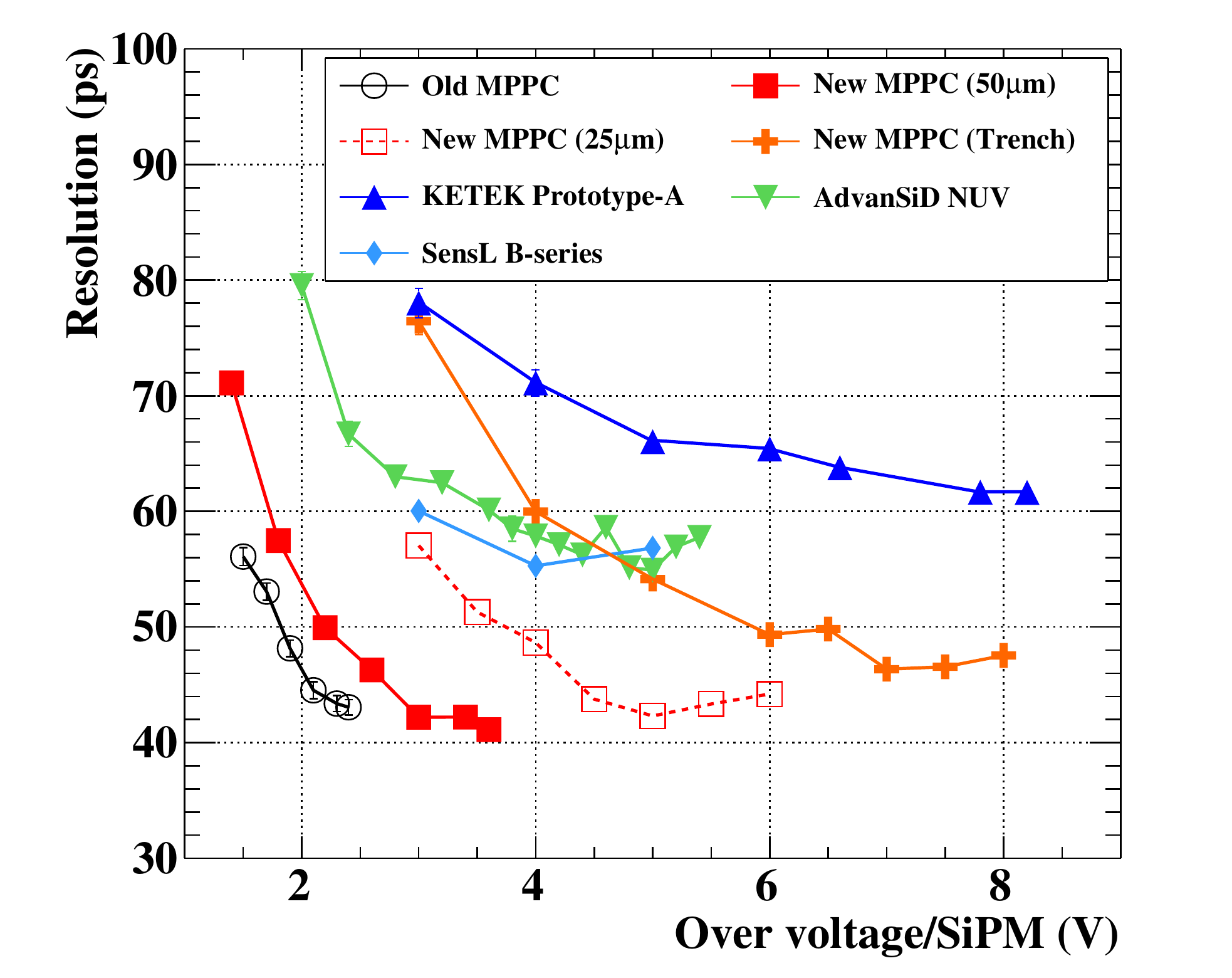}
        }
    \end{center}
    \caption{Summary of the SiPMs comparison. \emph{a)} Dark Count. \emph{b)} Cross Talk Probability \emph{c)} PDE \emph{d)} Time Resolution. All results are given as a function of the applied over-voltage. 
     }%
   \label{fig:subfigures}
\end{figure}

\paragraph{Dark noise and cross-talk}
The noise level of the device is evaluated by looking at the waveforms acquired by a random trigger. The dark count rate is calculated from the probability of observing zero photo-electron P(N$_{\mbox{p.e.}}$= 0) in a fixed time window. The result is shown in figure \ref{fig:darkcount}  as a function of the applied over-voltage.
\\
The cross-talk probability is calculated from the P(N$_{\mbox{p.e.}}\ge 2$)/P(N$_{\mbox{p.e.}}\ge 1$) ratio including a correction for the accidental coincidence of dark pulses. The result is shown in figure \ref{fig:crosstalk}. The cross-talk probability almost linearly increases with the over-voltage. The standard-type SiPMs, namely SiPMs without a trench structure, turned out to have worse performance with respect to the trench type whose improved structure strongly reduces the noise level. Anyway, the typical energy release in a pixel should guarantee an adequate signal-to-noise ratio also for those SiPMs with higher dark count and cross-talk rates.

\paragraph{PDE}
The PDE for Near UltraViolet (NUV) light is measured with a LED whose wavelength (370$-$410 nm) approximately matches the scintillator emission peak. The LED intensity is adjusted in such a way that the average number of observed photo-electrons ranges between 0.5 and 1.0. The relative PDE is then calculated from P(N$_{\mbox{p.e.}}$= 0) in accordance with Poisson statistics, and thus the measured PDE value does not include the effect of cross-talk nor after-pulsing. The result is shown in figure \ref{fig:pde}. The highest PDE is obtained with Hamamatsu S12572 model, with $50\, \mu m$ pitch cell. A more detailed description of noise and PDE studies can be found in \cite{bib:IEEEYusuke}.

\paragraph{Breakdown voltage versus temperature dependence}

\begin{figure}[ht!]
    \begin{center}
      \includegraphics[width=0.7\textwidth]{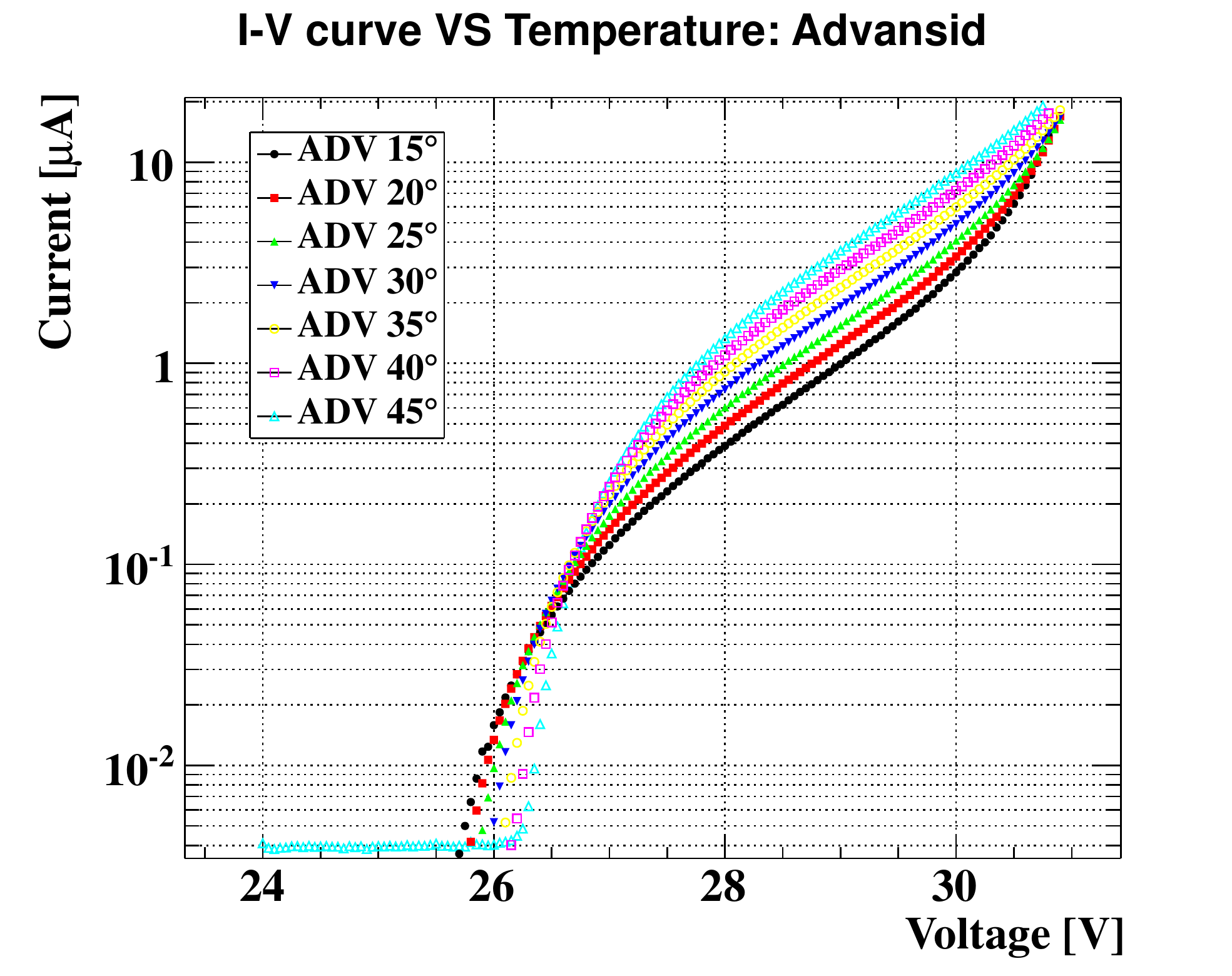}
    \end{center}
    \caption{I-V curves for different temperatures acquired with Advansid NUV SiPM.}
    \label{fig:BDcurve}
\end{figure}

The BreakDown voltage (BD) versus temperature dependence has been measured by plotting the I-V characteristic 
of each SiPM at different temperatures (see figure \ref{fig:BDcurve}) in the range $20^\circ\div 45^\circ$, 
resulting in a linear dependence,

\paragraph{Time resolution}
The basic setup for the timing resolution measurement is the same described above. A scintillator pixel with size $60\times30\times5$ mm$^{3}$ is read-out on each side by an array of 3 SiPMs connected in series. SiPMs are coupled to the pixel with optical grease.  A 35 ns coaxial cable (7.5 m) transports signals to amplifiers, simulating the final experimental conditions. Counters are excited by using a $^{90}$Sr $\beta$-source, providing electrons with 2.2 MeV endpoint energy. An external Reference Counter (RC) made of a small piece of scintillator (BC422, size: $5\times5\times5$ mm$^{3}$) coupled to a Hamamatsu S10362-33-050C SiPM is used for triggering purposes. The timing is extracted by applying a software constant fraction discrimination on the recorded waveform with discriminating fraction in the range $5\div10\%$ depending on SiPM model. 
\\
The time resolution of the system is evaluated as the width of the distribution $\Delta T = T_{RC}-\left( T_{0}+T_{1}\right)/2$, being $T_{RC}$ and $T_{i}$ the time measured by the reference counter and each SiPM array respectively. The RC resolution is evaluated $\sigma(T_{RC}) =30$ ps and subtracted. The summary of the results is shown in figure \ref{fig:timereso} as a function of the applied over-voltage. 

\subsection{Scintillator comparison}\label{sec:scintcomp}
Three types of ultra fast plastic scintillator from Saint-Gobain Crystals, BC418, BC420 and 
BC422, were tested. The main characteristics of each scintillator are summarised in table 
\ref{tab:scint}, where also the characteristics of the BC404 are listed. The test was 
performed using $60\times30\times5$ mm$^{3}$ pixels. The best resolution is obtained 
with BC422, the one with the fastest rise time.\\
\begin{table}[h]
\caption{Summary of the properties of plastic scintillators tested. Measured time resolutions on a $60\times30\times5$ mm$^{3}$ sample are also listed.}
\label{tab:scint}
\smallskip
\centering
\footnotesize
\begin{tabular}{|lcccc|}
\hline
Properties&BC404&BC418&BC420&BC422\\
\hline
Light Yield (\% Anthracene)&68&67&64&55\\
\hline
Rise Time (ns)&0.7&0.5&0.5&0.35\\
\hline
Decay time (ns) &1.8&1.4&1.5&1.6\\
\hline
Wavelength peak (nm)&408&391&391&370\\
\hline
Attenuation length (cm)&140&100&110&8\\
\hline
Measured resolution (ps) &-&48&51&43\\
\hline
\end{tabular}
\end{table}
%
%
%
\section{Beam test}
In order to test the detector in experimental conditions similar to the final one and check the multiple hit scheme, a small prototype was built and tested at the Beam Test Facility (BTF) at the INFN Laboratori Nazionali di Frascati \cite{bib:BTF}. The BTF beam can be tuned in such a way to provide electrons with energy similar to the MEG signal ($48$ MeV in our test) with average bunch multiplicity lower than 1. We decided to test counters equipped with both Hamamatsu and Advansid SiPMs, the ones with the best trade off between time resolution and temperature dependence.


\subsection{Setup}\label{setupBTF}

We prepared two sets of pixel prototypes with BC418 scintillator, with $90\times40\times5$ mm$^{3}$ sizes, equipped with Hamamatsu S12572-050C(X) (8 counters) and Advansid NUV (6 counters) SiPMs.\\
The pixels, wrapped with 3M Radiant Mirror Film,
are mounted on a moving stage that controls the movement in the plane perpendicular
to the beam. The whole system 
is mounted on an optical bench enclosed in a shielded black box. The same reference 
counter described in section \ref{sec:sipmcomp} is placed along the beam trajectory in front 
of the pixels. A lead glass calorimeter is placed behind the pixels for beam 
monitoring. The whole system is aligned to the beam line by using a 
laser tracker. Signals from SiPMs are fed into six DRS4 evaluation boards and sampled at 2.5 Gs/s. 

\subsection{Data analysis}\label{sec:anal}
\begin{figure}[ht]
   \centering
   \includegraphics[width=0.7\textwidth]{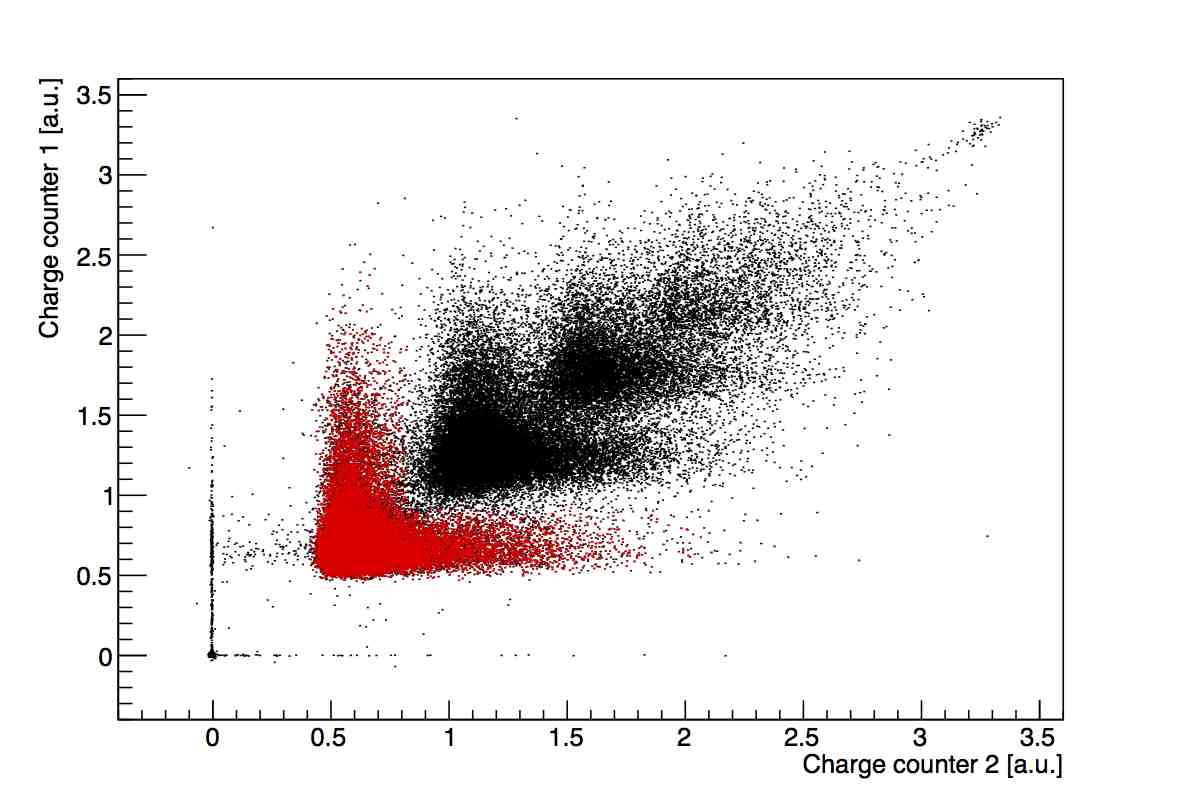}
   \caption{Charge distribution of the first couple of pixels. The selected events (single electron bunch) are marked in red.}
   \label{fig:Q1vsQ2}
\end{figure}

\paragraph{Charge analysis} Events are selected by cutting on the charge distribution of the first two pixels. An example of distribution is shown in figure \ref{fig:Q1vsQ2}, where the bunch multiplicity is clearly visible. 
Moreover, we applied also a cut on the reference counter charge spectrum, by selecting the events around the Landau peak of the charge distribution.

The timing resolution is then evaluated by taking the width of the $\Delta T$ distribution, defined 
in two different ways:
\begin{eqnarray}
\label{eq:DT1}
\Delta T(N) &=& T_{\mbox{RC}} - \frac{1}{N}\sum^{N}_{i=1} T_{i} \,,
\\
\label{eq:DT2}
\Delta T(N) &=& \frac{1}{\sqrt{2}}\left[\frac{1}{N/2}\sum^{N/2}_{j=1} T_{a_j}-\frac{1}{N/2}\sum^{N/2}_{i=1} T_{b_i}\right] \,,
\end{eqnarray}
where $T_{\mbox{RC}}$ and $T_{l}$ is the time measured by the reference counter and by the pixels respectively,
with $a_j$ and $b_i$ running on even and odd indices respectively. 
In formula \ref{eq:DT2} the sum is made over two different subgroups of pixels. In both cases, we can evaluate the timing resolution as a function of the number of hits used in the time averaging.

\paragraph{DRS calibrations}
Dedicated runs were taken to evaluate the contribution from the electronics jitter. It was found to be 18.7 $\pm 1.2$ ps and 16.2 $\pm 1.2$ ps for pixels whose arrays are read-out by the same or different boards, respectively. The former contribution is higher 
because the jitters from channels on the same baord are fully correlated. 

\subsection{Results}

\begin{figure}[htc] 
\centering
\includegraphics[width=.7\textwidth]{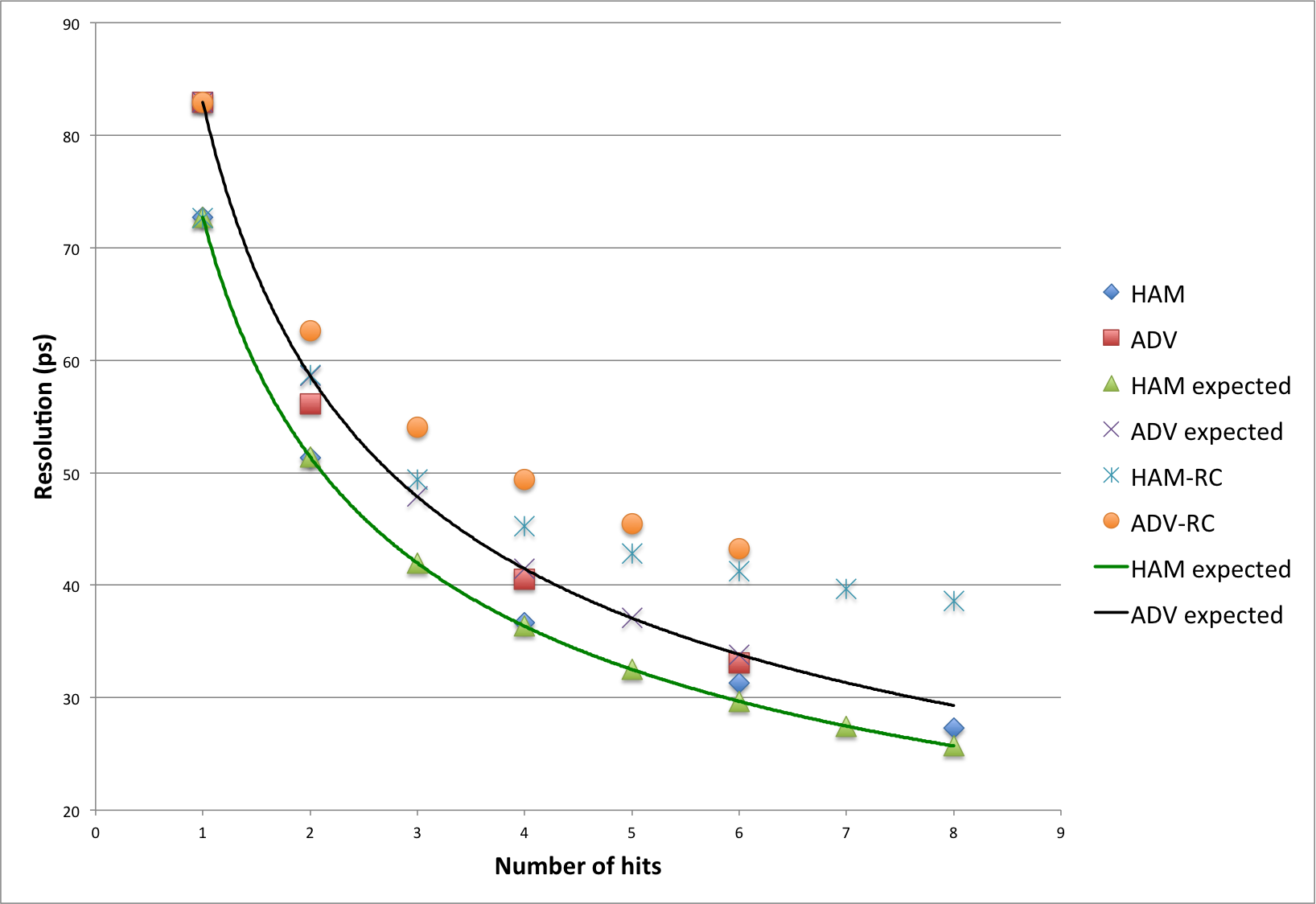}
\caption{Summary of the measured resolutions versus the number of hits: HAM (Hamamatsu) and ADV (Advansid) 
relying on the comparison of different pixel sets and HAM-RC and ADV-RC relying on the Reference Counter (RC).
The resolutions following the expected $1/\sqrt{N_{hit}}$ behaviour are also shown.
}
\label{fig:multihit}
\end{figure}

We checked the multiple hit scheme, relying on the approach in Eq.~\ref{eq:DT1} as in section \ref{sec:sipmcomp}
studying the time resolution versus the number of hits. 
The contribution from the electronics, described in section \ref{sec:anal} is taken into account. 
The RC resolution, which was checked with dedicated
runs and found to be $\sigma(T_{RC})=27$ ps is also subtracted.
As expected, the best result is obtained with the largest number of hits, with
$\sigma(\Delta T) < 30$ ps.
Preliminary resolutions are summarised in figure \ref{fig:multihit}, compared with the expected 
$1/\sqrt{N_{\mbox{hit}}}$ behaviour, which is also shown. An average $N_{\mbox{hit}} \sim 6.6$ is expected
in the experiment.

\section{Conclusions}

We presented the R\&D work on the upgrade of the Timing Counter for the MEG II experiment. 
The basic concepts of the new design, namely the good time resolution achievable with 
small scintillator counters read-out by SiPMs and the improvement of the overall time 
resolution by averaging the time measurements over multiple hits has been tested. 
Optimising the choice among different types of SiPM and scintillators leads to obtain 
extremely good time resolution with a single counter down to
$\sigma(\Delta T)\sim 43$ ps. A beam test performed at the 
Beam Test Facility in Frascati proved experimentally the multiple hit scheme. 
Analysis is still ongoing, a prelimiary resolution $\sigma(\Delta T)<30$ ps with 
eight counters is measured.

\acknowledgments

The authors thank the Beam Test Facility crew, the mechanical and electronics workshops at INFN Section of Genova and the Paul Scherrer Institute detector group for their valuable help.





\end{document}